# PLANNING CURRICULAR PROPOSALS ON SOUND AND MUSIC WITH PROSPECTIVE SECONDARY-SCHOOL TEACHERS

Erica Bisesi & Marisa Michelini

ABSTRACT
Sound is a preferred context to build foundations on wave phenomena, one of the most important disciplinary referents in physics. It is also one of the best-set frameworks to achieve transversality, overcoming scholastic level and activating emotional aspects which are naturally connected with every day life, as well as with music and perception. Looking at sound and music by a transversal perspective – a border-line approach between science and art, is the adopted statement for a teaching proposal using meta-cognition as a strategy in scientific education. This work analyzes curricular proposals on musical acoustics, planned by prospective secondary-school teachers in the framework of a Formative Intervention Module answering the expectation of making more effective teaching scientific subjects by improving creative capabilities, as well as leading to build logical and scientific categorizations able to consciously discipline artistic activity in music students. With this aim, a particular emphasis is given to those concepts – like sound parameters and structural elements of a musical piece, which are best fitted to be addressed on a transversal perspective, involving simultaneously physics, psychophysics and music.

## INTRODUCTION

*Teaching is the highest form of understanding*
Aristotle

We live in a world of sounds. We hear sounds every time a bird sings, listening to our hearth beating or the wind blowing outside. Discourse on sound lies on perceptual attributes of the auditory system, as well as on scientific reasoning about its origin and propagation. In the realm of physics, sound offers a very suitable context to build foundations on wave phenomena, one of the two models – besides particle approach – able to describe transfer of energy and momentum. All these aspects contribute to make sound a very important topic at any level physics training, justifying a dedicated course on this subject in the context of a Formative Intervention Module for physics education of secondary school teachers.

Although a topic not far from our everyday experience, a lot of factors contribute to make understanding the physics of sound lacking and difficult for students at any scholar level (Driver, 1994; Duit, 2007; Linder, 1989). First of all, students are not aware of some essential ideas regarding sound generation (sources) and propagation (wave nature of the signal, physical properties of waves, interaction with medium) and tend to resort to some abstractions to represent sound features with mental models (Menchen, 2005). Specifically, many authors recently investigated students' models of sound propagation, identifying *entity and hybrid models* as the main alternatives to *the scientifically accepted wave model* (Hrepic, 1998, 2004; Hrepic, Zollman and Rebello, 2002; Linder, 1993;

Maurines, 1998; Wittmann, 1998, 2001; Wittmann, Steinberg and Redish, 1999). In particular, according with the "entity" model, sound is a self-standing entity, different from the medium, and propagating through it (Hrepic, Zollman and Rebello, 2002). We observe that arising of such a conflict between two pictures – wave and particle-like models of sound, does directly link to one of the most troublesome problems underlying knowledge structures in modern physics: coupling between sources and medium in effective field theories. Moreover, a large number of difficulties and misconceptions are directly connected with the interdisciplinary nature of sound; in particular – an observation derived also by personal teaching experience, students can hardly understand the difference between concepts of *mechanical perturbation* and *physiological sensation,* and they are not aware of the fact that duration, loudness, pitch and timbre are interdependent properties (Merino, 1998b). More in general, a lot of blunders are made when mixing concepts and ideas belonging to disciplinary spheres of different hierarchical level, like physics, psychophysics, neurophysiology and psychology of music. In the following, we will refer to this trap as the *interdisciplinary swindle.*
Besides researches on learning problems, many studies show that students describe the phenomenon of sound following two pictures: a *microscopic perspective,* where sound in an entity that is carried by individual molecules and is transferred from one molecule to another through a medium, and the *macroscopic picture,* which threats sound as a bounded substance in the form of some travelling pattern (Hrepic, 2004; Linder and Erickson, 1989) (*micro-macro swindle).*

The awareness of giving a good preparation to future teachers led to the proposal for the Formative Intervention Module (MIF) for secondary school teachers outlined below. The Module has been devised in the light of the research done on learning processes in physics (Bosio and al., 1997; Duit, 2006) and on teacher training (Michelini, 2003; Michelini and Pugliese, 1999), also with the aid of multimedia materials (Loria and al., 1981; Mathelitsch, 2008; Michelini and al., 2001; MPTL, 2008; Pugliese and al., 1999). It has been characterized by the integration of disciplinary elements and planning addressed to professional tasks (Day and al., 1990; Eraut, 1994; Mäntylä, 2006), providing project products illustrated and analyzed below.

**CONTEXT**

The research we report deals with teacher training and aims at designing strategies to prepare teachers to target main learning problems on sound. The context was that of a biennial after Master Specialization School for Secondary Teaching (SSIS) at Udine University – Italy[1]. Students were 25 prospective teachers with Degrees in Mathematics (13), Physics (5), Astronomy (1) and Engineering (6).

**RESEARCH QUESTIONS AND METHODS**

Much of research about students' difficulties would be barely fruitless if efforts to improve students' understanding of scientific concepts would be not carried out. Drawing from the educational paradigm of action research, such efforts join two aspects: *i)* research activity and *ii)* design of teaching interventions which are innovative with respect to traditional approaches, because they address the same disciplinary content under different – and somewhat new – viewpoints, approaches and methodologies, or because new topics are introduced in the curriculum (Méheut and Psillos (2004);

---

[1] The SSIS is a biannual university school for pre-service teacher formation. In Italy, in the last ten years, graduates who wish to dedicate themselves to teaching have the opportunity to acquire necessary methodological, educational and psycho-pedagogic skills, with a special attention to laboratory activities. The study plan is divided into a common area of dedicated teachings of Education Science, as well as in specialized areas intended for training-disciplinary teaching into relative classes. Didactical plant is modular. Each academic year is divided into semesters; each module is worth 3 credits and takes place in 24 hours of teaching activity. Additional 30 credits are matured after apprenticeship, where students-teachers apply what has been planned in laboratories. Positive conclusion of the course gains teaching qualification for secondary school.

Testa, 2008). One of the most up-to-date issues in teacher education concerns possible contributions that research in physics education can provide to. The essential claim is that researchers have not to focus on learning issues *solely* by the perspective of students – neglecting the viewpoint of teachers, but the latter have to be *re-introduced* inside the learning process (Schulman, 1992; Testa, 2008). According with "pedagogical content knowledge" (PCK), to make topics comprehensible, any teacher has to have at hand a set of alternative forms of representation – powerful analogies, illustrations, examples, explanations and demonstrations – deriving both from research and in the wisdom of practice (Schulman, 1992). Therefore, the PCK essentially refers to the ability of a teacher to transform the disciplinary content in a form accessible to students and gives, at least in the first approximation, a measure of the effectiveness of a teacher. Research provides evidence that – by means of this approach – prospective teachers may become more conscious about their knowledge of the contents to teach in classroom, and more aware of their own role in teaching and learning processes.

The paradigm of PCK has been the central issue in developing the training course referred hereby, exploiting an integrated approach centred on disciplinary and pedagogical knowledge, in order to make future teachers able to choose the best way to offer the content to students. Within this picture, our educational strategy was designed to call attention of prospective teachers on the following aspects:

**Explicit elements:** *i)* pedagogical content knowledge which were considered**,** recovered and managed throughout the project; *ii)* learning problems and conceptual nodes known from research in physics education which set up project reference to be accounted for; *iii)* transversal elements, belonging to non-physical – i.e. perceptual or musical domains, to be included in curricular planning. Moreover, in order to promote prospective teachers' awareness about the effectiveness of developed tools, educational strategy was based on research results on the role of the laboratory and the effectiveness of modelling methodologies and technologies in teaching and learning processes.

**Implicit elements:** as the aim of our educational proposal was to help prospective teachers to increase understanding of sound in secondary school, we mainly focused on those aspects which are swindling with respect to general learning process – in this case crossing points between different conceptual domains or schemes. Following the scheme outlined above, main tasks addressed in the proposed pattern involve:
- overcoming difficulties set by the general tendency to adopt local *mental models,* common sense reasoning or partial points of view, i.e. antithesis between wave versus particle models of sound propagation;
- *interdisciplinary swindle:* turning points among different disciplines and levels involved in acoustical phenomena;
- *micro-macro swindle:* crossing and complementariness of microscopic and macroscopic structure of the system.

In the light of this epistemological setup, adopted working method was characterized by an interactive approach based on: *i)* teaching the main disciplinary elements grounding wave and sound foundations (Formative Intervention Module (MIF), see below); *ii)* creation of discussion groups aimed at addressing all elements listed above under pedagogical, learning problems-oriented and transversal point of views; *iii)* project production, consisting in designing curricular proposals by students-teachers, based on the educational pattern adopted in the course.

**FORMATIVE INTERVENTION MODULE**

The course comprised 10 lectures, organized in terms of a critical overview of the basic ideas on sound theory, and was structured in a modular way throughout the following steps:
- A. *disciplinary elements analysis:* wave production, propagation, transmission and reception processes; influence of the medium on wave shape and speed; wave description and graphics visualization in terms of physical quantities; difference between pulses and waves; superposition of waves; free and stationary waves; difference between concepts of wave and sound; difference between sound and noise; sound description in terms of both physical and perceptual attributes; sound path, from the source to the auditory system and to the brain; difference between sound and music; musical intervals and scales; theory of consonance; tempo and rhythm, 'horizontal' and 'vertical' components in sound and music; integration of rational and emotional aspects in sound and music;

B. *curricular aspects analysis:* examples of instructional paths on central physical concepts issues (Aiello and al., 1997; Fazio and al., 2006; Mazzega and Michelini, 1993; Sperandeo-Mineo and al., 2005);

C. *learning analysis by single elements:* critical discussion on main conceptual and learning nodes on waves and sound, as drawn by analysis of specialized literature: wave propagation (Hrepic, 1998, 2004, Hrepic, Zollman and Rebello, 2002; Wittmann, 1999, 2001); role of medium (Fazio and al., 2006; Wittmann, 1999, 2001); superposition of waves (Wittmann, 1999, 2001); resonance (Hrepic, Zollman and Rebello, 2002);

D. *analysis by mental models or learning swindles:* critical discussion on the conceptual bound between description of oscillating movement in terms of elementary systems – like springs or simple pendulum, and wave motion grounding sound production (Hrepic, 1998, 2004, Hrepic, Zollman and Rebello, 2002; Linder, 1993; Maurines, 1998; Wittmann, 1998, 2001; Wittmann, Steinberg and Redish, 1999); *micro-macro* swindle (Hrepic, 2004; Linder and Erickson, 1989); *interdisciplinary* swindle (Merino, 1998b);

E. *environmental learning analysis:* focus on previous knowledge and expected results; feedback with real world experience;

F. *transversal elements analysis:* relationship among different physical aspects; linking scientific disciplines together; emphasis on the connection between physical wave features – like superior partials and envelopes, and psychophysical attributes of sound; scientific-humanistic bound; mathematical foundation of musical subjects (intervals, scales, chords harmony, theory of consonance, rhythm categorization); historical development of main ideas on waves and sound.

As a leit-motiv underlying our didactical proposal, we claim that the two theoretical basis of brain competencies (Mac Lean, 1949) – the theory of stratified evolution of human brain and the theory of specialization of brain hemispheres, can be powerfully accounted for the multi-dimensional picture where music arises from the cooperation between horizontal (temporal and melodic) and vertical (dynamical and harmonic) model structures. Specifically, horizontal dynamic describes volume distribution inside subsequent grouped sounds, strongly impressing listener sensitivity with an immediate impact on emotions; differently, vertical dynamics indicates volume ratios inside single sound events, distributing any voice inside sound space (Altenmüller, 2005). According with this picture, music listening contemporarily activates both analytical-spatial and cortical-emotional complementarities: looking at sound and music in such a transversal perspective – a border-line approach between science and art, has been the adopted statement for a teaching proposal using meta-cognition as a strategy in scientific education.

To increase effectiveness of proposed approach, a wider use of new technology in the classroom was resorted to; main instruments were:
- a particular care on graphics representation of sound, aimed at both powerfully describing wave features in real situations – like superposition, stationary waves in musical instruments, superior partials, beats and combined sounds, and giving a picture on how different aspects of sound are linked together (i.e., spectrum, sonogram, harmonic table vibration figures);
- interactive numerical tools, provided both by web applets and dedicated software to simulate sound properties and production processes;
- audiovisual presentation of topics and examples (sounds, images, movies, interactive tools);
- scientific technology;
- musical instruments;
- evaluation tools (income and outcome tests, student cards).

The course concluded asking prospective teachers to draft a curricular proposal on sound, according with the following road-map:
- by using offered topics and discussed learning problems as a resource, individuating an intervention path proposal for secondary school;
- by involving the largest number of sound aspects and spacing among the highest degree of conceptual perspectives, create a transversal education nature in the path;
- by using ideas, materials and strategies coherent with the proposed pattern, carry out educational tools (frontal lectures, exercises or laboratory activity) to be coupled to theoretical treatment.

**DATA ANALYSIS**

Aiming at understanding if our material influenced the quality of teaching planning – i.e. thematic integration on different proposed levels, patterns coherence, formalization level and attack angles – we analyzed all reports in the light of both emphasizing focus on addressed disciplinary, learning and transversal aspects and assessing the level of coherence and purpose featured by proposed curricular projects. In order to better characterize any pattern, we estimated richness and quality of educational materials, with reference to adopted elements and strategies.

**Methods**
We performed a qualitative analysis following empirical research methods where research questions are dominant with respect to class analysis (ESERA, 2006) of the following relevant aspects:
- disciplinary content elements;
- pedagogical-learning elements;
- transversal elements.

Hence, each grid component has been estimated by a rate or a qualitative mark. At a first stage, we evaluated completeness of disciplinary grid by grouping subjects according with the number of elements. All percentages are calculated counting the elements present in each report, with respect with those in the steps A) – F) of the MIF, as listed above:
1. *disciplinary content elements:*
    *a)* no mention;
    *b)* less than 33% of the whole presented;
    *c)* between 33% and 50%;
    *d)* between 50% and 66%;
    *e)* above 66%.

Moreover, learning and transversal elements were weighted and rated according to the following scheme:
2. *pedagogical approach, learning strategies and tools:*
    *0:* no mention;
    *1:* only a list of topics;
    *2:* proposals to address conceptual knots or learning problems;
    *3:* activity sequences and operative proposals oriented to overcome learning difficulties;
3. *transversal elements:*
    *0:* no mention;
    *1:* without internal coherence;
    *2:* with internal coherence;
    *3:* with resort to transversality to overcome learning problems.

Finally, we analyzed both kinds and quality of proposed materials and tools. We performed a first qualitative analysis by quantifying elements plenty (formulas, graphics, software, audiovisuals, multimedia, scientific technology, musical instruments, evaluation tests and cards, other) and reusing context (frontal lecture, exercises, or laboratory experiments):

4. *materials and tools* (we considered only organic materials, supporting stated proposals):
   *a)* no mention, only talkative treatment of the subject;
   *b)* less than 33% of the whole presented in the course;
   *c)* between 33% and 50%;
   *d)* between 50% and 66%;
   *e)* above 66%.

At a subsequent stage, we built exclusive categories related to the quality of the path, according with the following requirements:
- explicit resort to educational tools to overcome learning problems;
- integration with interdisciplinary elements;
- methodological rigours, i.e. explicit discussion and/or planning of strategies and methods with respect to single activities;
- focus on motivation.

**RESULTS AND DISCUSSION**

Distribution of students-teachers, according with the number of disciplinary elements they have considered, is shown in Figure 1. As we can infer from the diagram, disciplinary elements are uniformly distributed between low (33% – 50%), intermediate (50% – 66%) and high responses (above 66%). This result gives very encouraging information on general people involvement, proving flexibility and appreciation of our approach. Anyway, the most important implication is *that distributions of elements in learning and transversal classes are not biased by disciplinary elements richness, but only depend on the way they have been framed and rearranged into the learning context.*

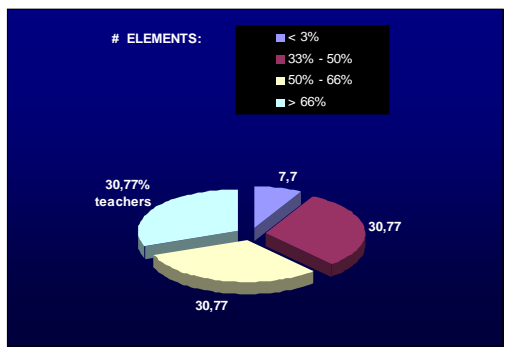

Figure 1. Distribution of students-teachers according with the number of disciplinary elements considered in their works. Data set are shown in the panel.

Detailed data analysis gives us indication of a wider use of fundamental wave properties, exhaustively discussed both by words and graphic pictures, as well as awareness of the role of medium on wave propagation. Another successful topic is the connection between stationary waves and superior partials, framed into musical perspective involving intervals, scales and musical instruments' behaviour. Difference between physical and perceptual parameters is clear, but this is not the case for conceptual changes between pulse and wave, wave and sound, and sound and music. Treatment of sound way, from external source to human brain, is dealt with a discrete level, whereas a certain sensitivity to emotional aspects activated by music listening seems not sufficiently characterize different cognitive and disciplinary levels.

Data analysis on pedagogical and transversal elements is shown in Figure 2. In this case, average weighted rates show a good understanding and contextualization of disciplinary nodes and learning problems *(upper left)*, as well as a copious use of transversal links *(lower).* By comparing upper left and lower panels in Figure 2, we infer that elements of the two areas are strongly correlated: *contextualization and overcome of learning elements seems to track coherence and learning-oriented* *purpose of transversal processes and schemes.* A confirmation of this statement is provided by the comparison between partial rates of single students: the *correlation coefficient* between the two set of measures in upper left and lower panels is *r = 0.36,* meaning a corresponding correlation probability above *75%.* We remark hereby that the MIF gave no reference to any explicit expectation on possible connection between learning context and transversality, so that *results outlined so far do be ascribed to a picture implicitly and spontaneously arising during knowledge building process.*

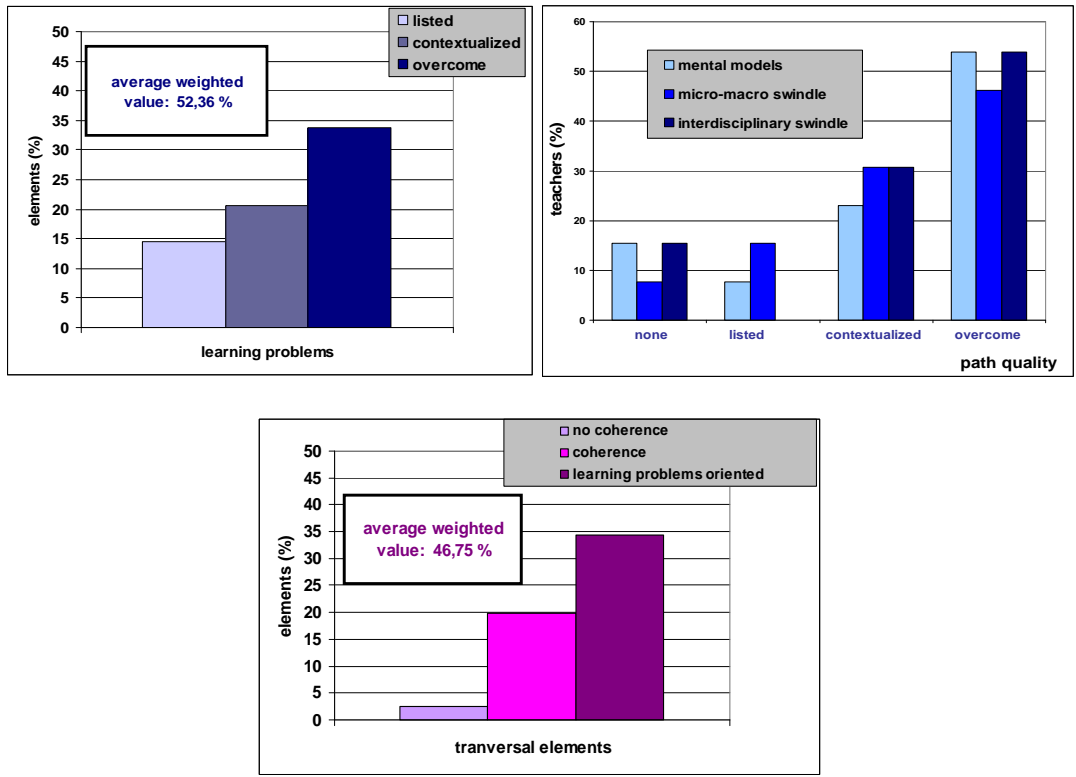

Figure 2. Learning and transversal elements data. *Upper left:* distribution of learning aspects, according with their contextualization and operative proposals for overcoming. *Upper right:* implicit aspects related to learning elements, weighted by quality of relative paths (percentages referred to teachers' sample). *Lower:* distribution of transversal elements, focusing on the coherence and purpose to association with learning problems. Average weighted values are calculated according with prescriptions outlined in the text above.

Data analysis allowed to point out *implicit aspects* invoked into research questions, namely *resort to mental models, micro-macro and interdisciplinary swindles.* As we can infer from upper right panel in Figure 2, all these aspects have been addressed with a high degree of contextualization and operative proposals oriented to overcome learning difficulties. Additionally, students-teachers were more careful with disciplinary prerequisites than with expected results, and show a positive feedback with external world. They better felt at their ease with problem solving connected with reasoning than with single conceptual nodes, in agreement with previous studies certifying effectiveness of learning through conceptual change with respect to single concepts acquirement and meaning chains (Hewson, 1981; Posner and al., 1982). Connections among both single physical aspects and different disciplines are well explored; nevertheless, this target is reached only when tranversality is grounded on mathematical foundations – like in the historical picture of interval consonance theory.

Figure 3, 4 and 5 respectively acquaint with material usage and proposed didactical strategies (percentages referred to teachers' sample). As we can easily see, teachers are very flexible in resorting to user-friendly innovative tools – namely interactive procedures and applets management. Moreover, many projects did explicitly refer to experimental laboratory instrumentation, as well as involved

playing musical instrument in classroom. Nevertheless, traditional teaching means – i.e. mathematical or graphical description of phenomena – seem to be preferred both in frontal lectures and in formulation of empirical questions during laboratory activity. In spite of the large number of proposed examples and tools for computer simulation – a very powerful and sophisticated mean systematic musicology can avail itself, both in teaching and expressive performance – resort to this kind of strategy is very poor and reticent. Finally, evaluation instruments – i.e. income and outcome tests or experiment student cards – are taken in very little consideration at all.

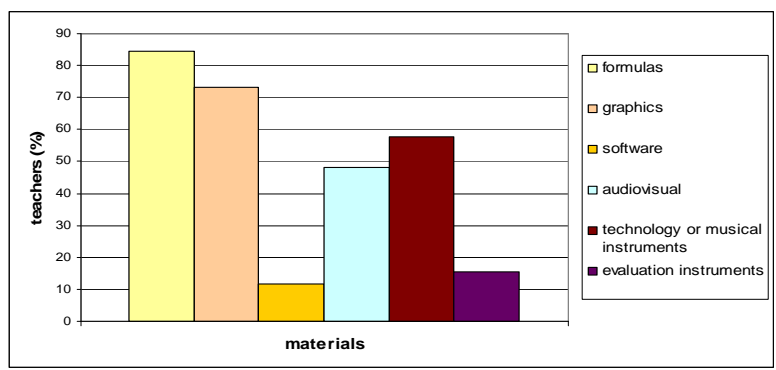

Figure 3.  Educational tools resorted to by students-teachers in their curricular proposals.

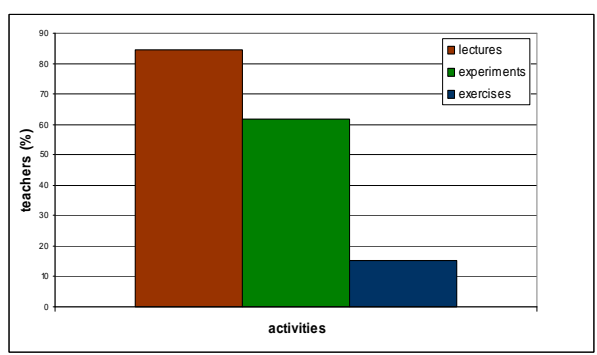

Figure 4.  Didactic activities distribution.

Materials shown in Figure 3 were presented in the framework of mainly interactive frontal lessons and laboratory activity (Figure 4). This activity context has been further analyzed according with methodological rigours and attention given to learning questions or transversality, as well as to psychological motivations by secondary school students. Following the scheme outlined above, we built 20 exclusive evaluation categories, whose representative instances are drawn in Figure 5. *The most outstanding result is the constant usage of interdisciplinary elements in all the educational material proposed, as well as – once more – a deep relationship between interdisciplinary aspects and main conceptual nodes or learning swindles.* Teachers' homework is also well structured under a methodological point of view, and motivational aspects addressed to pupils' involvement are taken into consideration.

## CONCLUSIONS AND IMPLICATIONS

We analyzed curricular proposals for teaching sound in secondary school, drawn by 25 prospective teachers at the end of a MIF aimed at designing strategies to prepare teachers to target main learning problems on waves and sound. The MIF – carried out in the context of a biennial after Master

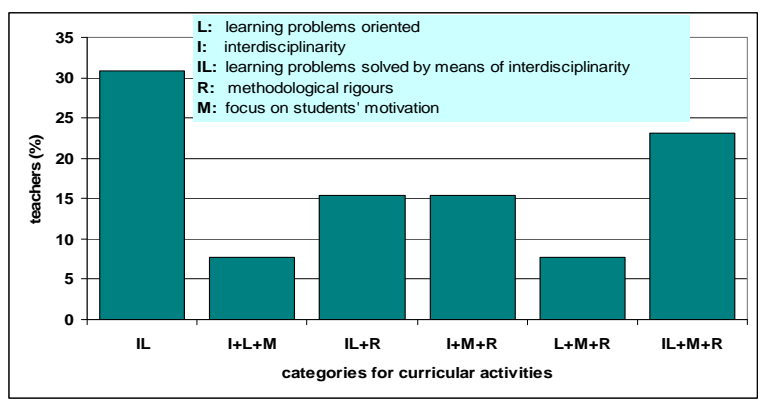

Figure 5. Representative exclusive categories related to the quality of the path, illustrating curricular activities of Figure 4.

Specialization School for Secondary Teaching (SSIS) at Udine University, Italy – comprised a critical overview of the basic ideas on sound theory, discussions on the main conceptual nodes known from research in physics education, and a transversal approach involving elements belonging to non-physical, i.e. perceptual and musical domains. Our research strategy was centred on the quantity of content elements considered in curricula, as well as on the purpose to lead teachers inside learning process, transforming disciplinary content in a form accessible by students by means of alternative forms of representation (PCK paradigm). The course concluded asking prospective teachers to propose original curricular paths, by using offered topics and discussed learning problems as a resource and by creating a transversal education nature in the approach. We find that content elements are uniformly distributed between low (33% – 50%), intermediate (50% – 66%) and high responses (above 66%), proving that distributions of learning and transversal elements are not biased by content elements richness, but only depend on the way they have been framed and rearranged into the learning context. Data analysis on pedagogical elements shows a good contextualization and resort to proposals for solution of learning problems (~ 55%). Additionally, main implicit tasks addressed in the proposed pattern – i.e. resort to mental models, micro-macro and interdisciplinary swindles – emerge with a high degree of contextualization and operative proposals oriented to overcome learning difficulties. As regards the ability to transform resource materials, transversal elements are generally coherently addressed and we measured a high correlation factor (> 75%) between operative proposals oriented at overcoming learning difficulties and resort to transversal aspects. Whereas quite anchored at traditional teaching means – mathematical and graphical description of phenomena, developed in the framework of frontal lessons, teachers were very flexible in resorting to user-friendly – multimedia and interactive – innovative tools, as well as to laboratory experimental instrumentation and musical instruments. Teachers' homework is also well structured under a methodological point of view (explicit discussion and/or planning of strategies and methods with respect to single activities) and motivational aspects addressed to pupils' involvement are taken into deep consideration. All these general results – supported also by lack of any intentional indication about targets during the course – give indication of that a transversal approach to waves, sound and music may provide a valid support to overcome general learning problems associated with wrong mental models and conceptual bounds.

Erica Bisesi
Post-doc researcher
Learning in Physics Group
Department of Physics
University of Udine
Via delle Scienze, 208
33100 – Udine (UD)
Italy
Email: ericabisesi@yahoo.it, bisesi@fisica.uniud.it

Marisa Michelini
Full Professor
Learning in Physics Group
Department of Physics
University of Udine
Via delle Scienze, 208
33100 – Udine (UD)
Italy
Email: michelini@fisica.uniud.it